\documentclass[aip,jcp,reprint,twocolumn]{revtex4-1}
\usepackage{bm}
\usepackage{amsmath}
\newcommand{\abs}[1]{\left| #1 \right|}
\newcommand{\grad}{\bm{\nabla}}
\newcommand{\upDelta}{\mathop{}\!\Delta}

\usepackage[utf8]{inputenc}
\usepackage[T1]{fontenc}
\usepackage[group-separator={,}]{siunitx}
\usepackage{enumitem}

\usepackage{graphicx}
\graphicspath{{fig/}}

\usepackage{multirow}
\usepackage{booktabs}
\newcommand{\tworow}[1]{\multirow{2}{*}[-0.5\dimexpr \aboverulesep + \belowrulesep + \cmidrulewidth]{#1}}

\usepackage{cleveref}
\Crefname{figure}{Fig.}{Figs.}

\begin{document}
\title{A robust and memory-efficient transition state search method for complex energy landscapes}
\date{\today}

\author{Samuel J. Avis}
\author{Jack R. Panter}
\email[]{j.r.panter@durham.ac.uk}
\author{Halim Kusumaatmaja}
\email[]{halim.kusumaatmaja@durham.ac.uk}
\affiliation{Department of Physics, Durham University, South Road, Durham DH1 3LE, UK}

\begin{abstract}
  Locating transition states is crucial for investigating transition mechanisms in wide-ranging phenomena, from atomistic to macroscale systems.
  Existing methods, however, can struggle in problems with a large number of degrees of freedom, on-the-fly adaptive remeshing and coarse-graining, and energy landscapes that are locally flat or discontinuous.
  To resolve these challenges, we introduce a new double-ended method, the Binary-Image Transition State Search (BITSS).
  It uses just two states that converge to the transition state, resulting in a fast, flexible, and memory-efficient method.
  We also show it is more robust compared to existing bracketing methods that use only two states.
  We demonstrate its versatility by applying BITSS to three very different classes of problems: Lennard-Jones clusters, shell buckling, and multiphase phase-field models.
\end{abstract}

\maketitle

\section{Introduction}
Transition states are central to the description of reconfiguration mechanisms for systems in chemistry, condensed matter physics, and engineering.
Historically, many computational methods for locating transition states have grown from an atomistic or particulate perspective.
These have proven to be important tools for understanding, for example, protein folding \cite{Bryngelson1995,Onuchic1997}, biological and industrial catalysis \cite{Boehr2006,Kerns2015,Guo2018a}, quantum tunnelling \cite{Richardson2016,Vaillant2019}, crystallisation \cite{Richard2018}, and cluster formation \cite{Wales1998,Wales2012}.

More recently, it is increasingly being recognised that transition states are useful in mesoscale or macroscale systems.
Here, the minimum energy barriers provide important lower bounds to the energy input required for transitions to occur.
This has been used to understand failure in structural engineering applications \cite{Panter2019,Hutchinson2018}, for the development of super liquid-repellent surfaces \cite{Zhang2014,Panter2019b,Amabili2017}, and investigating locomotion through complex terrain for robotics \cite{Othayoth2020}.
Moreover, it is becoming desirable to tailor elastic deformation transitions to enable technologies such as advanced deployable structures \cite{Filipov2015,Zhai2018}, mechanical sensors and actuators \cite{Bertoldi2017,Truby2016,Chi2022,Bonfanti2020}, and energy absorbers \cite{Shan2015,Giri2021} to name but a few.

Transition state search methods generally fall into two categories, single- and double-ended methods.
Single-ended methods are initialised at a single state and attempt to climb to a nearby saddle point.
Examples include eigenvector following \cite{Cerjan1981}, the dimer method \cite{Heyden2005,Kastner2008,Zhang2016}, and climbing image methods \cite{E2007,Ren2013}.
Double-ended methods can be further subdivided into two groups.
The first utilise a chain of states between two minima which is then minimised to provide an estimate for the full transition pathway in addition to the transition state.
Examples are the string method \cite{E2002,E2007} and doubly-nudged elastic band (DNEB) \cite{Trygubenko2004}.
These methods require an appropriate initial interpolation, which can sometimes be challenging to obtain \cite{Wales2012a}.
The second group are bracketing methods, which involve two states converging to the transition state from either side.
These include the Dewar-Healy-Stewart (DHS) algorithm \cite{Dewar1984}, ridge method \cite{Ionova1993}, the step and slide method \cite{Miron2001}, and the double-ended surface walking method \cite{Zhang2013}.

A large range of landscapes, however, prove challenging or impossible to explore via these methods.
One key problem arises from the push towards larger and more complex systems \cite{Trefethen2013,Shalf2020,Alexander2020}, resulting in the need to develop algorithms that are more computationally and memory efficient, and can incorporate optimisation strategies such as on-the-fly adaptive remeshing and coarse-graining.
These typically involve changing the resolution or discretisation of the systems to focus the computational time on important regions, such as using a higher resolution mesh in regions of high stress in finite element simulations \cite{Lee1994}.
However, chain-of-states methods involve a coupling between the configurations of each state, and so there is an issue if they have different discretisations and numbers of degrees of freedom.
Meanwhile, single-ended methods can be inefficient because they are not well suited for identifying specific pathways and can spend a large amount of time searching for undesired transition states.
Another major challenge in studying complex energy landscapes relates to the presence of locally flat or discontinuous regions, such as when considering patchy \cite{McMullen2018,Nguemaha2018,Chen2018b} and hard-body \cite{Richard2018,Santra2018} interactions in atomistic simulations, systems of polymer chains \cite{Mokkonen2016}, or collision constraints for macroscopic objects \cite{Wriggers2006}.
Flat zero-modes in the landscape pose issues for single-ended search methods and current bracketing methods that rely only upon local information.
Specialist treatment can sometimes be used such as in the case of global rotation and translation \cite{Page1988}, but they are thwarted by local zero-modes.
Finally, current methods cannot typically be applied in the case of discontinuous potentials, or if the gradient is prohibitively expensive to compute, because continuous, differentiable optimisation functions are required.

In this work we introduce a new double-ended bracketing method, the Binary-Image Transition State Search (BITSS).
Using a range of different applications, we demonstrate that it successfully addresses each of the above challenges.
In addition, we show that BITSS is superior compared to existing bracketing methods, allowing us to access the transition states when other methods fail.

\section{BITSS method}
\begin{figure}[tb]
  \includegraphics[width=\linewidth]{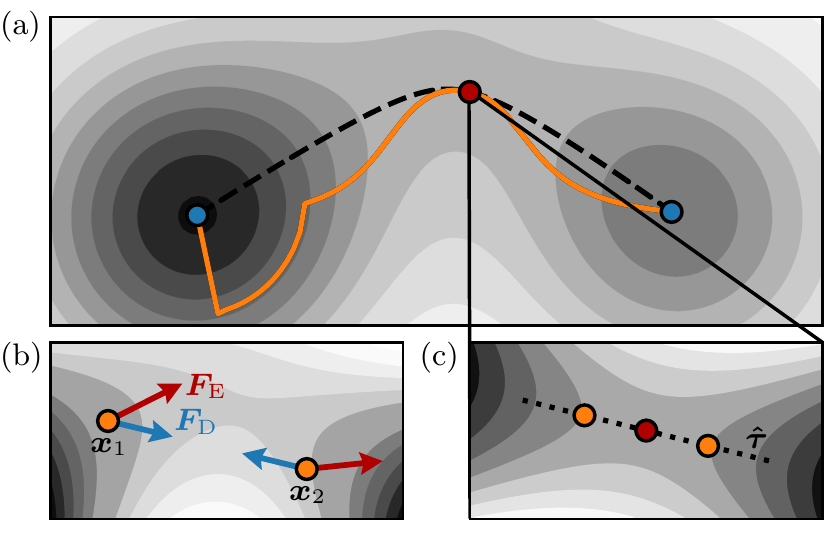}
  \caption{\label{fig:toy2d}
    Schematics of the BITSS method on a simple 2D potential with two minima.
    (a) The orange line shows the trajectories of the two states from the minima (blue) to the transition states (red) under the BITSS method.
        The minimum energy pathway is shown by the dashed line.
    (b) A snapshot of the BITSS minimisation showing the driving forces on each state due to the energy constraint, $\bm{F}_\mathrm{E}$, and distance constraint, $\bm{F}_\mathrm{D}$, with $E_1 < E_2$ and $d(\bm{x}_1,\bm{x}_2) < d_i$.
    (c) The final configuration of the BITSS method showing the two states in orange, the transition state in red, and the negative curvature eigenvector, $\bm{\hat{\tau}}$.
  }
\end{figure}

The method begins by first initialising the states, $\bm{x}_1$ and $\bm{x}_2$, in the basins of attraction of different local minima, such as the two blue spots in the 2d potential in \cref{fig:toy2d}a.
These can be set to the minima, but this is not a necessary requirement.
The energies of these two states are then minimised, while constraining their separation.
This is iteratively reduced to zero, such that, at iteration $i$, their separation is
\begin{equation}\label{eq:diteration}
  d_i = (1 - f) d_{i-1},
\end{equation}
with $d_0$ taking the value of the separation between the two initial states.
A reduction factor of $f = 0.5$ is successful for most applications, but this can be made smaller to ensure that the states do not slide off the ridge between the two basins of attraction.
Different metrics may be used to compute this distance, although in this work we simply use the Euclidean distance,
\begin{equation}
  d(\bm{x}_1, \bm{x}_2) = \sqrt{\sum_i (x_{1,i} - x_{2,i})^2}.
\end{equation}
To further ensure that neither state is pulled over the ridge, a secondary constraint enforces equal energies for the two states.
Using this strategy, the two states will meet at the lowest point on the ridge, the transition state.

The two constraints are implemented using energy penalty terms, which result in driving forces on the two states if the constraints are not met, such as in \cref{fig:toy2d}b.
Including these energy penalty terms gives the total BITSS energy for the pair of states,
\begin{multline}\label{eq:bitss}
  E_\mathrm{BITSS}(\bm{x}_1, \bm{x}_2) = E_1 + E_2
    + \kappa_e \left( E_1 - E_2 \right) ^2 \\
    + \kappa_d \left( d(\bm{x}_1, \bm{x}_2) - d_i \right) ^2,
\end{multline}
where $E_1$ and $E_2$ are the single-state energies, and $\kappa_e$ and $\kappa_d$ parametrise the strengths of the energy and distance constraints.

In this work the L-BFGS algorithm is chosen to minimise this energy, owing to its fast convergence and low memory requirement for large numbers of degrees of freedom \cite{Liu1989}.
However, any other minimisation method can be used instead.

To ensure that the transition state is located successfully, the constraint strengths $\kappa_d$ and $\kappa_e$ are updated as the algorithm proceeds using information from the system.
These are set such that the driving forces due to the constraints and single-state energies are of similar size.
This prevents the constraints from dominating the underlying potential or causing large jumps that make a state pass over the ridge.
This results in the following equations (see Supplementary Note I for the derivation),
\begin{gather}
  \kappa_e = \frac {\alpha} {2 E_\mathrm{B}},
  \label{eq:ke}
  \\
  \kappa_d = \max \left(
    \frac {\sqrt{\abs{\grad E_1}^2 + \abs{\grad E_2}^2}} {2\sqrt{2} \beta d_i} \; , \;
    \frac{E_\mathrm{B}}{\beta d_i^2} \right),
  \label{eq:kd}
\end{gather}
where $\grad E_1$ and $\grad E_2$ are the gradients of the energies of the two states, and $\alpha$ and $\beta$ are parameters with recommended values of $\alpha = 10$ and $\beta = 0.1$.
Here, $E_\mathrm{B}$ is an estimation for the current energy barrier, evaluated using the difference between the highest energy along a linear interpolation between the two states and the average energy of the two states.
These constraints are initially calculated at the start of each minimisation, and regularly recalculated throughout (once per 100 iterations is used in this work).

In practice, when numerically minimising, the states will jump about slightly which can result in large gradients perpendicular to the optimal movement direction.
To reduce this effect, the gradients used in \cref{eq:kd} are projected in the direction of the separation between the two states:
\begin{equation}
  \abs{\grad E_n} \approx \frac {\abs{(\bm{x}_1 - \bm{x}_2) \cdot \grad E_n}} {\abs{\bm{x}_1 - \bm{x}_2}}.
\end{equation}

In summary, the method involves iteratively performing the following three steps:
\begin{enumerate}[noitemsep,nolistsep]
  \item Reduce the constrained separation, according to \cref{eq:diteration}.
  \item Minimise the potential of the pair of states, \cref{eq:bitss}.
  \item Recompute the constraint coefficients, $\kappa_e$ and $\kappa_d$, at regular intervals using \cref{eq:ke,eq:kd}.
\end{enumerate}
This process is completed once a suitable convergence criterion is reached.
This can either be based upon the separation between the states, the size of gradient at the midpoint between them, or the change in the position of the midpoint.

Using the BITSS approach, the typical trajectories of the states are demonstrated for a simple 2D potential in \cref{fig:toy2d}a.
Initially, the lower energy state jumps up to satisfy the equal energy constraint and then moves to minimise the separation without increasing its energy.
Then, the two states converge directly towards one another, before being deflected towards the saddle in the ridge.
Consequently, if there are multiple possible pathways between two states, BITSS will be biased towards identifying those that are more direct or with lower energy.
Furthermore, the final two states are positioned either side of the transition state in the direction of the negative curvature eigenvector, $\bm{\hat{\tau}}$ (\cref{fig:toy2d}c).
So, BITSS automatically identifies the `reactive mode' and associated eigenvalue in addition to the transition state.
Once the transition state has been identified, it is possible to find the full minimum energy pathway by tracing the trajectory of downhill minimisations from the two final states, which are either side of the saddle.

In the event that there are intermediate stable states, there will be a chain of multiple transition states between the two minima.
In this case, the equal-energy constraint will not prevent the states from passing over the lower energy transition states, so BITSS should converge to the transition state with the highest energy.
This enables the identification of the overall energy barrier, providing estimates about the overall ease of the transition, or the rate for chemical processes.
However, as demonstrated in Supplementary Note II, if multiple transition states have very similar energies then a smaller distance reduction factor, $f$, may be necessary to ensure that it does indeed converge to the highest transition state.
Furthermore, if all of transition states or the full pathway are desired, BITSS can be continually repeated from one of the minima downhill from the located transition state and one of the initial minima until the initial minima are piecewise connected by a full pathway.

\section{Results and Discussion}
\subsection{Comparison with other bracketing methods}
\begin{figure}[htb]
  \includegraphics[width=\linewidth]{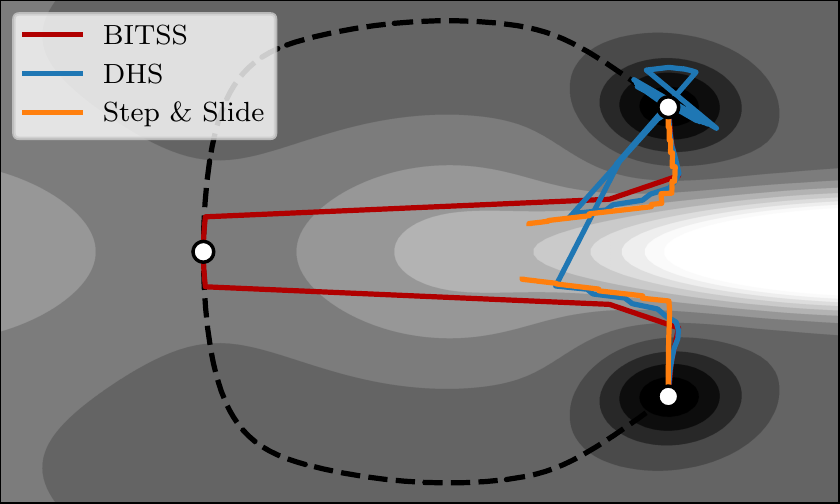}
  \caption{\label{fig:bracketing}
    Trajectories of the bracketing methods on a hooked potential with a single saddle point.
    The equation for this potential is provided in \cref{sec:2dpotentials}.
    The minimum energy pathway is shown by the dashed line.
  }
\end{figure}
The BITSS potential in \cref{eq:bitss} and iterative steps above offer key advantages over existing bracketing methods that also use two states to locate the transition state.
For instance, in the ridge method \cite{Ionova1993}, the two images are initially chosen to bracket the largest energy point on an interpolated path between the two endpoints.
However, this is not guaranteed to be on the ridge containing the transition state, and specialist methods are required to avoid high-energy local maxima, or when the initial path contains multiple candidate maxima.
In another example, the double ended surface walking method \cite{Zhang2013} requires Gaussian bias potentials to be added at each iteration to force two dimers to climb uphill in the landscape.
For high numbers of degrees of freedom and many iterations, this becomes very computationally expensive.

The two methods most similar to BITSS are the DHS \cite{Dewar1984} and step and slide \cite{Miron2001} methods.
In the step and slide method, the separation between two images is minimised while their energy is fixed (iteratively increasing the energy up to the transition state).
Conversely in the DHS method, the energy of an image is minimised while the image separation is fixed (iteratively decreasing the separation and changing the frozen image up to the transition state).
To illustrate how BITSS is superior compared to these methods, we consider the hooked 2d potential in \cref{fig:bracketing}.
For this potential the energies of the images ascend higher than that of the transition state and consequently both of these methods fail to converge to the saddle point regardless of the parameters that are used.
The step and slide method fails in this situation because it always expects that the energy of the two states is below the saddle point if they have not converged, so it has no means of descending down the ridge.
For DHS, the images reach a certain point at which one state can pass over the ridge by minimising its energy.
At this point DHS will fail even if the distance is reduced very slowly.
In contrast, the BITSS method is successful for this potential.
This is because the combination of distance and energy constraints allow BITSS to approach a transition state from both below and above (by sliding down a ridge).

Furthermore, using both an energy and distance constraint with BITSS provides improved efficiency over these methods which each use just one of the constraints.
In the case of DHS, fixing one state in place and optimising the other means that the amount that the separation is reduced must be much smaller than BITSS to ensure that it does not pass over the ridge.
Meanwhile, in step and slide, it is difficult to obtain a reasonable energy increment when the two states are far from the transition state, leading to a larger than necessary number of iterations.
Also, restricting the minimisation to a constant energy surface can result in a considerably more complex method, as the states must be constantly projected back onto this surface.

\subsection{Comparison with chain-of-states methods}
In contrast to BITSS, chain-of-states methods do not typically find transition states directly.
Instead, they are designed to find the full pathway (or an approximation thereof), and a secondary method can then be used to refine to the transition state.
As we will demonstrate, this strategy is successful for simple, linear pathways, but faces two key challenges when the pathway is highly non-linear.
Firstly, for such complex pathways, a large number of states are required to sufficiently approximate the minimum energy pathway.
The second is that choosing a suitable initial interpolation can be problematic to achieve.
BITSS can be advantageous in both these regards, as only two states are evolved, regardless of the pathway complexity, and no initial interpolation is required.

Here we compare the speeds of convergence of BITSS to two widely used approaches for finding transition states that employ chain-of-states methods: climbing image nudged elastic band (CINEB) \cite{Henkelman2000a}, and DNEB with hybrid eigenvector following (DNEB-HEVF) \cite{Cerjan1981}.
The core of these methods involves minimising the total energy of a chain of states, connected by elastic springs to keep them equally spaced along the transition pathway.
We fix the two end-points at the minima, so the number of states that are minimised is two fewer than the number of states in the chain.
CINEB modifies the method by altering the behaviour of the state with the highest energy.
The direction of minimisation on this state is inverted along the pathway direction, effectively converting the saddle point into a local minimum.
Alternatively, DNEB-HEVF involves first minimising the chain of states until a convergence criterion is met, and then performing hybrid eigenvector following from the highest energy state, moving uphill along the smallest eigenvector of the Hessian until it reaches the transition state.
For completeness, we also combine hybrid eigenvector following with BITSS and include the results in the convergence comparison.
Additional implementation details for these methods are included in \cref{sec:HEVFdetails}.

It is also possible to use the string method with a climbing image \cite{E2007} or eigenvector following \cite{Zimmerman2013}; although, in this case, the results are expected to be similar to the nudged elastic band methods.
We note that our aim in this section is to observe how the BITSS method behaves for different systems, rather than providing a comprehensive comparison of the currently available methods, which has been performed in other works \cite{Koslover2007,Sheppard2008}.

\begin{figure}[tb]
  \includegraphics{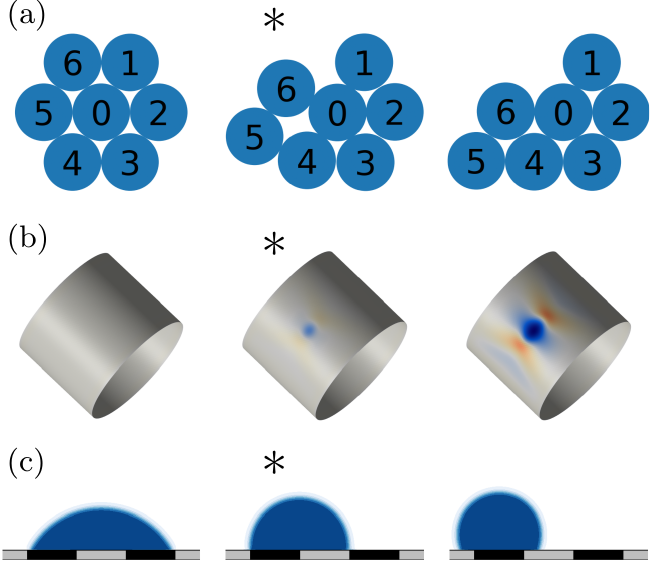}
  \caption{\label{fig:testsystems}
    The three systems used for comparison with chain-of-states methods.
    These are: (a) a Lennard-Jones seven-particle cluster, (b) cylindrical shell buckling, and (c) wetting of a chemically-striped surface.
    The configurations shown correspond to the two minimum energy states and the transition state, marked by an asterisk.
  }
\end{figure}

Three diverse systems are used for this comparison, exhibiting a broad range of energy landscapes.
The first system is a two-dimensional, seven-particle cluster, interacting via a Lennard-Jones pair potential.
This is a frequently used test system for studying transition rates \cite{Wales2002,Passerone2001}.
Here, the 14 degrees of freedom are the particle coordinates.
The characteristic transition shown in \cref{fig:testsystems}a sees a particle rearrangement between two close-packed clusters.

The second system is an elastic cylindrical shell, modelled by a triangulated mesh of nodes, which interact via extensional and angular springs.
The \num{35400} degrees of freedom are the node coordinates in three-dimensional space.
The characteristic transition in \cref{fig:testsystems}b shows the formation of a stable dimple from an initially unbuckled cylinder.
This transition is essential to capture and predict mechanical failure under strain \cite{Panter2019,Virot2017}.

\begin{table*}[tb]
  \begin{ruledtabular}
  \begin{tabular}{c cc cccc cccc}
    \tworow{System} & \tworow{BITSS} &
      \tworow{\begin{tabular}[c]{@{}c@{}}BITSS-\\HEVF\end{tabular}} &
        \multicolumn{4}{c}{DNEB-HEVF} & \multicolumn{4}{c}{CINEB} \\
    \cmidrule(lr){4-7}\cmidrule(lr){8-11}
             &             &             & 3           & 5           & 10          & 20          & 3   & 5          & 10          & 20          \\
    \hline                                                                                                                                             
    LJ-7     & 148         & 144         & 135         & 138         & 208         & 361         & 30  & 153        & 2000        & 1692        \\
    Buckling & \num{12866} & \num{14446} & \num{73694} & \num{7315}  & \num{6242}  & \num{12094} & --- & \num{8352} & \num{19360} & \num{77904} \\
    Wetting  & \num{12100} & \num{17262} & \num{17840} & \num{17721} & \num{20221} & \num{48286} & --- & ---        & ---         & ---         \\
  \end{tabular}
  \end{ruledtabular}
  \caption{\label{tab:speed}
    Number of potential gradient calculations required to reach the transition state for the three comparison examples.
    The climbing image nudged elastic band (CINEB) and DNEB with hybrid eigenvector following (DNEB-HEVF) methods have been run for different numbers of images.
    Convergence is determined to be when the root-mean-square of the gradient at the estimate for the transition state is less than $10^{-4}$.
    The fields left blank indicate that the method has not converged to the correct transition state.
  }
\end{table*}

The final system involves a droplet situated on a chemically striped surface with both hydrophilic and hydrophobic regions.
Droplet transitions on patterned surfaces such as this are vital to understand as powerful bio-inspired liquid manipulation strategies  \cite{Kusumaatmaja2006,Brown2016}.
In the example shown in \cref{fig:testsystems}c, a droplet transitions from two hydrophilic patches to one patch.
Here, the system is represented by a diffuse-interface model, in which the \num{40000} degrees of freedom are the local fluid compositions at each site of the discretised domain.
The pathway for this example is highly non-linear in the coordinate space because each degree of freedom only varies when it is at the interface of the droplet.
As a result, the initial pathway for the chain-of-states methods cannot be a simple linear interpolation.
Instead, the position of a semi-circular droplet is interpolated between the two final positions.

The results for the three systems are shown in \cref{tab:speed}.
First, we note that for all three systems, using hybrid eigenvector following does not significantly improve the speed of BITSS.
Indeed, for buckling and especially wetting, HEVF is detrimental to performance.
Next, it is interesting to compare each method's performance between simple and complex pathways.
In contrast to the wetting transition's highly non-linear pathway, the pathways of the LJ-7 rearrangement and the buckling system can be simply tracked following a gradual variation in the order parameters.
For LJ-7, this is the translation of atoms 5 and 6, and for buckling, this is the radial displacement of the centre of the dimple \cite{Panter2019}.
For the simpler pathways, BITSS is generally slower, but for the complex pathways, BITSS is faster.
Moreover, we see that for the wetting example, CINEB does not converge to the transition state because the estimated tangent vector is highly inaccurate due to the non-linearity of the pathway.

For situations where memory is limited, it is important to minimise the number of images used.
However, efficiently finding the TS is challenging for both CINEB and DNEB-HEVF if too few images are used, as observed for the cylindrical buckling with three images.
BITSS, on the other hand, converges using only two images.

\subsection{Adaptive discretisation}\label{sec:adaptive}
\begin{figure}[tb]
  \includegraphics[width=\linewidth]{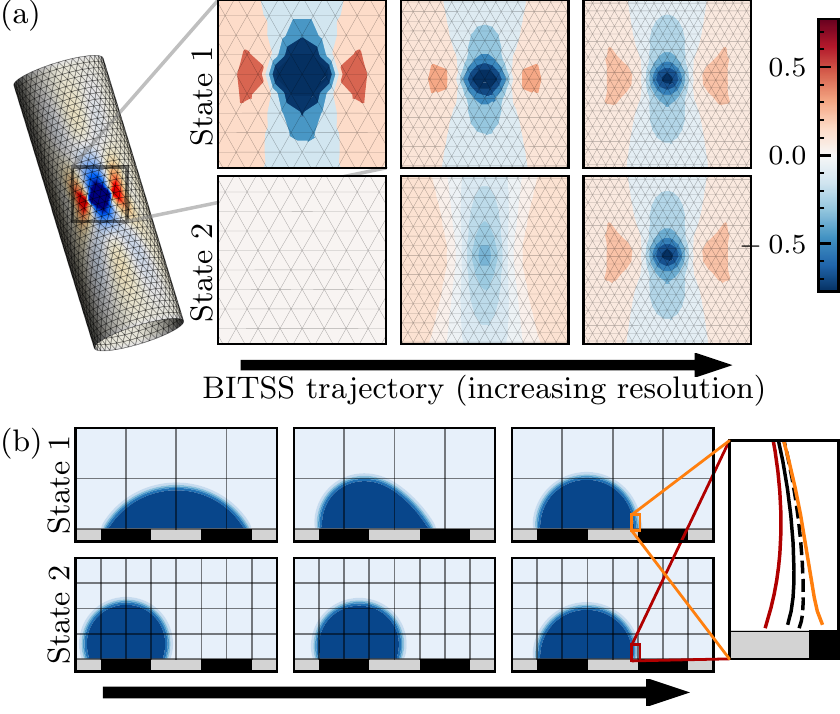}
  \caption{\label{fig:adaptivemesh}
    Demonstration of BITSS addressing the challenges associated with adaptive remeshing.
    (a) Snapshots of the BITSS method for the buckling of a cylinder with a changing mesh.
        The radial displacement relative to the unbuckled cylinder is shown, as well as the underlying triangular mesh.
    (b) Snapshots for the striped wetting example with different resolutions for the two states.
        Each grid cell denotes 50x50 lattice nodes.
        The zoomed axis shows the difference in the fluid interface between the two final states, as well as the approximated transition state (solid black line).
        This is compared to the transition state found using a high resolution (dashed line).
  }
\end{figure}

Adaptive remeshing and coarse-graining are widely used techniques that we can utilise to further increase the efficiency of BITSS.
These techniques cause issues for most existing double-ended methods because the coupled states may end up with different degrees of freedom.
However, in BITSS the only direct coupling is in the distance measure, $d(\bm{x}_1,\bm{x}_2)$, which is relatively easy to adapt.
Here we demonstrate the use of adaptive remeshing by considering two separate issues.

Firstly, we show in \cref{fig:adaptivemesh}a that BITSS is able to handle the discretisation adapting, and the number of degrees of freedom changing, as the method runs.
For this we use the cylindrical buckling example with the resolution increasing from 40 to 100 triangles around the cylinder, corresponding to an increase from \num{1760} to \num{11000} degrees of freedom.
This demonstrates that BITSS is able to converge to the transition state so long as the remeshing is not so significant as to shift a state into the basin of attraction of the other minimum.

In the second test, shown in \cref{fig:adaptivemesh}b, we demonstrate the use of different meshes for the two states in the striped wetting example.
In this case, the distance measure is adapted by interpolating one state onto the other mesh and computing the Euclidean distance.
However, for some applications a simpler measure may be sufficient, such as the difference between average values of the system.
Using this approach, BITSS is able to closely approach the transition state.
The precision of this convergence is now limited by the transition state energy differing slightly on each grid, but this effect will be reduced when using an adaptive method or a higher resolution.

\subsection{Complex landscapes}
\begin{figure}[tb]
  \includegraphics[width=\linewidth]{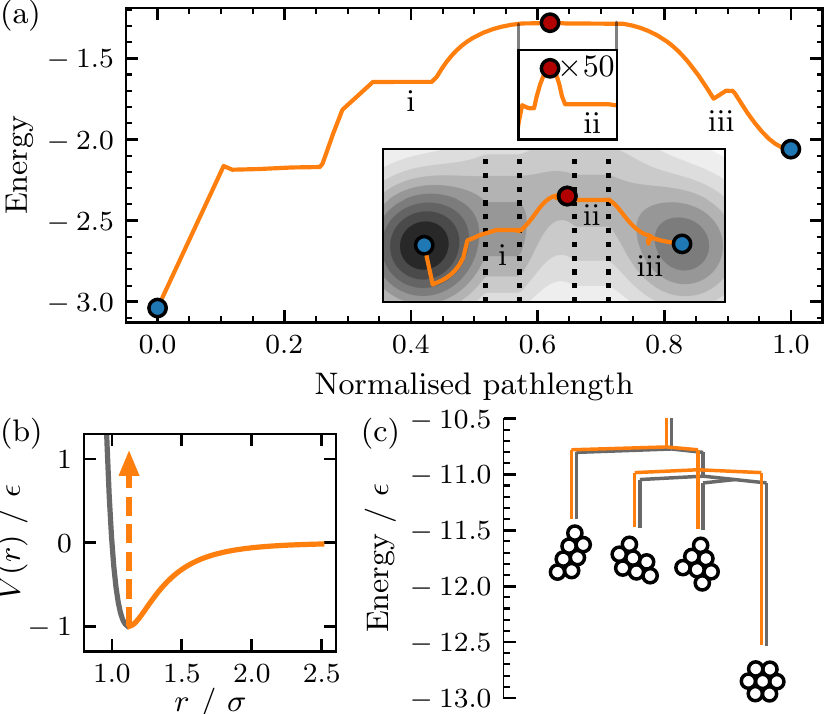}
  \caption{\label{fig:flatdiscontinuous}
    Demonstration of BITSS applied to flat and discontinuous potentials.
    (a) Energy profile of the BITSS pathway on a 2D potential with flat regions.
        Blue and red dots denote the minima and transition state, respectively.
        Points of interest are labelled by i--iii (see text).
        Top inset: A zoomed in view around the transition state.
        Bottom inset: The pathway taken, with the edges of the flat regions marked by dashed lines.
    (b) The discontinuous hard-core pair potential used in the seven-particle cluster (orange).
        The standard Lennard-Jones potential is also shown in grey.
    (c) Disconnectivity graphs of the energy landscapes for the seven-particle cluster with the two potentials.
        The two graphs are offset for visibility.
  }
\end{figure}

The final challenges we will address are those related to complex landscapes that prove challenging for previous algorithms.
The first is the presence of flat regions in the landscape.
\Cref{fig:flatdiscontinuous}a shows BITSS applied to a 2D landscape with two such regions (i \& ii) that are flat in the $x$-direction.
We see BITSS is able to successfully converge past these flat regions, even with one very close to the transition state (ii).
In these regions there are no driving forces due to the potential and the energy constraint, which use purely local information about the gradient.
However, the distance constraint continues to pull the states together, preventing them from getting stuck.
When only a single state has a zero-gradient mode then the other is likely to slide down the potential slightly (iii), but the two states still remain either side of the dividing ridge and so the result is unaffected.

An additional consideration is the case where the potential energy surface is flat at the top of the pathway.
There are two possibilities here, one is that the potential is flat in a direction perpendicular to the tangent of the pathway, such that the ridge is level.
In this case BITSS is unaffected and it will be able to converge to some point along the ridge.
An example of this is the free global rotation and translation of the Lennard-Jones cluster in \cref{fig:testsystems}a.
The other possibility is that the flat mode is in the direction of the pathway.
In this case there is no single transition state along the pathway, but instead a region.
BITSS would be ill suited in this situation because the equal-energy constraint would not prevent the images from passing over the saddle and falling to a minimum.

Finally, we investigate the application of BITSS to systems with undefined gradients, such as when the landscape is discontinuous.
To account for this, the equations for the coefficients must be adapted to not depend upon the gradients, and a gradient-free minimiser (simulated annealing) is used.
These changes are detailed in the methods section.
This has been tested using a 7-particle cluster with a hard-core Lennard-Jones pair-potential, shown in \cref{fig:flatdiscontinuous}b, which results in a discontinuous landscape.
Using the gradient-free approach, BITSS is able to successfully find the transition states, allowing us to plot the disconnectivity graph of the system, shown in \cref{fig:flatdiscontinuous}c.
Compared with the results for the standard Lennard-Jones cluster, the energies of the minima are largely unchanged, but the energies of the transition states are found to be slightly higher.
This indicates that the particles in the Lennard-Jones cluster cut the corner slightly as they transition, whereas this is not possible using the discontinuous potential, resulting in higher energies.
Despite this gradient-free method being feasible, it is worth noting that a gradient-based approach is significantly more efficient, and so should be preferred if possible.

\section{Conclusion}
Overall, we have developed the binary image transition state search (BITSS) algorithm for the efficient location of transition states in traditionally challenging landscapes.
This has distinct advantages for complex pathways owing to the lack of a required initial pathway estimate, as well as the identification of the transition state that provides the overall energy barrier in multi-step pathways.
From the speed analysis, we find that the combination of chain-of-states methods with single-ended transition state search methods provides good performance for near-linear pathways, such as for the Lennard-Jones cluster and cylindrical buckling.
However, for highly complex and non-linear pathways, as exhibited by the striped wetting example, BITSS is superior.
Indeed, the demonstrated speed and memory-efficiency will be key as we move towards studying larger and more complex systems using BITSS.

A second source of efficiency in the BITSS method comes from the ability to adaptively change the degrees of freedom as the algorithm proceeds.
We demonstrated how transition states could be found by both increasing the resolution upon convergence, and coupling systems with different discretisations.
The ease of coupling two copies of a system and adaptive remeshing, now leads to the possibility of incorporating BITSS into existing open-source optimisation methods, such as surface evolver \cite{Brakke1992} or finite element methods \cite{Kolev2021}, to provide important energy barrier functionality.

Finally, we showed how BITSS can be used to survey discontinuous energy landscapes, demonstrated for a system of attractive hard-core particles.
This opens up possibilities for studying a broad range of systems previously out of reach of conventional landscape methods, but where transition information is valuable.
These include systems with very short range interactions, such as in colloidal clusters, or hard contact forces, such as in the folding of elastic materials, or locomotion and environmental interaction in robotics.

The distance metric between the two BITSS images is interesting to analyse further.
One question that emerges is whether transition states can be located by coupling two images through a small number of collective properties, rather than the total distance between all degrees of freedom in the system.
A second question concerns landscapes with multiple competing pathways between states.
In such cases, it may be possible to access transition states different from the most direct one by using a biased distance metric.
A further investigation that is now open to pursue is when discontinuities in the landscape occur at `stationary points' (now properly referred to as critical points).
In this case, a transition state can no longer be defined by its Hessian eigenvalues, but instead is more broadly defined as a region of locally minimal energy that separates two basins of attraction to minima.
Overall, it will be interesting to explore how BITSS enables access to even more challenging landscapes, and those not yet amenable to traditional landscape exploration techniques.

\section*{Supplementary material}
See the supplementary material for derivations of the expressions of the constraint coefficients in \cref{eq:kd,eq:ke}, and the demonstration of the BITSS method applied to a path with multiple transition states.

\begin{acknowledgments}
  S.~J.~A. is supported by a studentship from the Engineering and Physical Sciences Research Council [Grant No. EP/R513039/1].
  H.~K. and J.~R.~P. acknowledge funding from the Engineering and Physical Sciences Research Council [Grant No. EP/V034154/1].
\end{acknowledgments}

\section*{Conflict of Interest}
The authors have no conflicts to disclose.

\section*{Author Contributions}
All authors conceived and planned the project.
S.~J.~A. performed the numerical experiments and analysed the data.
J.~R.~P. and H.~K. supervised the study.
S.~J.~A. wrote the paper, with review and input from and J.~R.~P. and H.~K.

\section*{Data availability}
The data that support the findings of this study are available from the corresponding author upon reasonable request.

\appendix
\section{BITSS: Changes for undefined gradients}
A couple of alterations to the method must be made to account for situations where the gradients are unknown.
First, the calculation of $\kappa_d$ in \cref{eq:kd} must be adapted to avoid the use of gradients.
This can be done by simply removing the first term and just using the second term in the equation.
Secondly, L-BFGS can no longer be used because it requires knowledge of the gradients.
We must instead use a minimiser that does not require a differentiable optimisation function, for which we use simulated annealing \cite{Kirkpatrick1983}.
This has a chance of randomly jumping one state over the dividing barrier, but we can reduce this probability by limiting the initial temperature and maximum random displacement.
We typically employ $T_0 = E_\mathrm{B} / 10$, and $d_\mathrm{max} = d(\bm{x}_1, \bm{x}_2) / 100$.

\section{Adaptive discretisation test details}
Here we provide the details for the interpolations and mapping involved in the two examples demonstrating the feasibility of using an adaptive discretisation method.
For the cylindrical buckling example with a changing mesh, the resolution is refined each time the separation between the two states has halved, and is performed at the end of each iteration of the BITSS method.
This involves the number of triangles around the circumference of the cylinder increasing along the sequence: $40 \rightarrow 60 \rightarrow 80 \rightarrow 100$; with the number of degrees of freedom increasing by: $\num{1760} \rightarrow \num{3960} \rightarrow \num{7040} \rightarrow \num{11000}$.
The positions of the nodes on the new grid, $\{\bm{n}_i\}$, are determined by linear interpolation from the previous grid $\{\bm{p}_i\}$, using the positions of the unbuckled meshes, $\{\bm{n'}_i\}$ and $\{\bm{p'}_i\}$.
For each node of the new grid, $\bm{n'}_i$, the triangle that contains it is first identified, which we will denote $\{\bm{p'}_1,\bm{p'}_2,\bm{p'}_3\}$, and the barycentric coordinates of the point are computed, $\{\lambda_1,\lambda_2,\lambda_3\}$.
The new position is then given by $\bm{n}_i = \lambda_1 \bm{p}_1 + \lambda_2 \bm{p}_2 + \lambda_3 \bm{p}_3$.

In the wetting example with different resolutions for the two states, the distance is obtained by first mapping the phase field from the higher resolution grid, $\{\phi_{k,l} | k,l \in \{0,1,\cdots,399\}\}$, to the low resolution grid, $\{\phi'_{i,j} | i,j \in \{0,1,\cdots,199\}\}$.
Because a square grid is used with a resolution ratio of two, the mapping involves averaging each 2x2 block to a single point:
\begin{equation}
  \phi'_{i,j} = \frac{1}{4} \left( \phi_{2i,2j} + \phi_{2i+1,2j} + \phi_{2i,2j+1} + \phi_{2i+1,2j+1} \right).
\end{equation}
Then the separation from the other state, $\{\widetilde{\phi}_{i,j}\}$, is computed using the 2-norm,
\begin{equation}
  d = \sqrt{\sum_{i,j} \left( \phi'_{i,j} - \widetilde{\phi}_{i,j} \right)^2}.
\end{equation}
Finally, the gradient of the distance with respect to each point must be mapped back to the higher-resolution grid, which is done by assigning a quarter of each component back to its the original four points:
\begin{equation}
  \frac{\partial d}{\partial \phi_{k,l}} =
    \frac{\partial \phi'_{i,j}}{\partial \phi_{k,l}} \frac{\partial d}{\partial \phi'_{i,j}} =
    \frac{1}{4} \frac{\partial d}{\partial \phi'_{i,j}} =
    \frac{1}{4} \frac{\phi'_{i,j} - \widetilde{\phi}_{i,j}}{d},
\end{equation}
where $k \in \{2i, 2i+1\}$, and $l \in \{2j, 2j+1\}$.

\section{Energy and gradient expressions for the example systems}
\subsection{2D potentials}\label{sec:2dpotentials}
The 2D potentials in \cref{fig:toy2d,fig:bracketing} use a sum of Gaussian potentials, \cref{eq:gaussianpeaks}, the parameters for which are provided in \cref{tab:toy2dparams,tab:bracketingparams}.
\begin{equation}
  E(x,y) = \sum_i a_i \exp\!\left( - \frac{(x-b_{x,i})^2}{c_{x,i}} - \frac{(y-b_{y,i})^2}{c_{y,i}} \right)
  \label{eq:gaussianpeaks}
\end{equation}

\begin{table}[ht]
  \begin{ruledtabular}
  \begin{tabular}{ccccc}
    $a$ & $b_x$ & $b_y$ & $c_x$ & $c_y$ \\
    \hline
    -3  & -1.4  &   0   &   1   &   1   \\
    -2  &  1.4  &   0   &   1   &   1   \\
    -1  &  0.07 &   1   &   1   &   1   \\
  \end{tabular}
  \end{ruledtabular}
  \caption{\label{tab:toy2dparams}
    Parameters for the Gaussians to produce the potential used in \cref{fig:toy2d}.
  }
\end{table}

\begin{table}[ht]
  \begin{ruledtabular}
  \begin{tabular}{ccccc}
    $a$  & $b_x$ & $b_y$ & $c_x$ & $c_y$ \\
    \hline
    -1   &   0   &   0   &  10   &  10   \\
     1   &   0   &   0   &   1   &   1   \\
     5   &   2   &   0   &   1   &  0.1  \\
    -1   &   1   &   1   &  0.1  &  0.1  \\
    -1   &   1   &  -1   &  0.1  &  0.1  \\
    0.01 &   0   &   0   &   1   &   1   \\
    0.5  &  -2   &   0   &   1   &   1   \\
  \end{tabular}
  \end{ruledtabular}
  \caption{\label{tab:bracketingparams}
    Parameters for the Gaussians to produce the potential used in \cref{fig:bracketing}.
  }
\end{table}

\subsection{Particle cluster system}
In the particle cluster example system, the Lennard-Jones potential is used for the interaction between each pair of particles.
Therefore, the potential and its gradient for each pair of particles are
\begin{gather}
  E = 4\epsilon \left[ \left(\frac{\sigma}{r}\right)^{12} - \left(\frac{\sigma}{r}\right)^6 \right], \\
  \frac{\partial E}{\partial \bm{x}_1} = -\frac{\partial E}{\partial \bm{x}_2} =
    \frac{24 \epsilon (\bm{x}_2 - \bm{x}_1)}{r^2} \left[ 2 \left(\frac{\sigma}{r}\right)^{12} - \left(\frac{\sigma}{r}\right)^6 \right],
\end{gather}
where $\bm{x}_1$ and $\bm{x}_2$ are the positions of the two particles, $r$ is their separation, $\epsilon$ is the interaction strength, and $\sigma$ is the particle radius.

\subsection{Cylindrical buckling system}
A 2D triangular mesh is used to model the cylindrical buckling system, with the ends of the cylinder fixed in place to apply an axial compression of 0.14\%.
The energy of the system is evaluated by treating all bonds in the mesh as an elastic spring to obtain the stretching energy, and all pairs of adjacent triangles to be connected by elastic hinges, providing the bending energy.
Their expressions are given by
\begin{equation}
  E = \sum_i k^\mathrm{S}_i (r_i - r^0_i)^2 + \sum_j k^\mathrm{B}_j [1 + \cos(\theta_j - \theta^0_j)].
\end{equation}
The first term is the stretching energy, where $r_i$ is the length, $r^0_i$ is the equilibrium length, and $k^\mathrm{S}_i$ is the stretching rigidity of bond $i$.
The second term provides the bending energy, where $\theta_j$ is the dihedral angle, $\theta^0_j$ is the equilibrium angle, and $k^\mathrm{B}_j$ is the bending rigidity of hinge $j$.

\begin{figure}[htb]
  \includegraphics{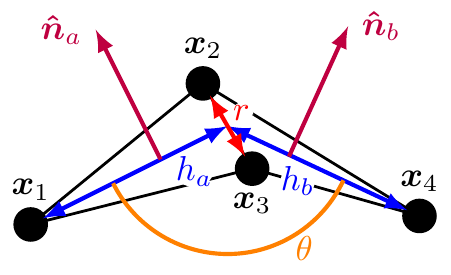}
  \caption{\label{fig:BAHschematic}
    Schematic of the bar and hinge model showing the relevant parameters for a single hinge element.
    $h_i$ and $\bm{\hat{n}}_i$ denote the height and unit normal of each triangle respectively.
  }
\end{figure}
The gradient of the energy can be obtained by individually considering the stretching and bending energies of a single bond and hinge.
For simplicity, we will ignore the index for the bond and hinge.
Using the variables shown in the schematic in \cref{fig:BAHschematic}, the gradient of the stretching energy of the bond between $\bm{x}_2$ and $\bm{x}_3$ is given by,
\begin{equation}
  \frac{\partial E^\mathrm{S}}{\partial \bm{x}_2} = - \frac{\partial E^\mathrm{S}}{\partial \bm{x}_3} =
    2 k^\mathrm{S} (r - r^0) (\bm{x_2} - \bm{x_3}).
\end{equation}
The gradients of the bending energy of the hinge are
\begin{align}
  \frac{\partial E^\mathrm{B}}{\partial \bm{x}_1} &= k^\mathrm{B} \sin(\theta - \theta^0) \frac{\bm{\hat{n}}_a}{h_a}, \\
  \frac{\partial E^\mathrm{B}}{\partial \bm{x}_2} &= - k^\mathrm{B} \sin(\theta - \theta^0) \left[\frac{\bm{\hat{n}}_a}{h_a} + \frac{\bm{\hat{n}}_a + \bm{\hat{n}}_b}{r}\right], \\
  \frac{\partial E^\mathrm{B}}{\partial \bm{x}_3} &= - k^\mathrm{B} \sin(\theta - \theta^0) \left[\frac{\bm{\hat{n}}_b}{h_b} + \frac{\bm{\hat{n}}_a + \bm{\hat{n}}_b}{r}\right], \\
  \frac{\partial E^\mathrm{B}}{\partial \bm{x}_4} &= k^\mathrm{B} \sin(\theta - \theta^0) \frac{\bm{\hat{n}}_b}{h_b}.
\end{align}

\subsection{Striped wetting system}
This is modelled on a 200x200 2D grid (and 400x400 in \cref{sec:adaptive}) using a phase-field model \cite{Panter2019b}, which has an order parameter, $\phi(\bm{r})$, representing the phase of the liquid ($\phi=-1$ for gas, $\phi=1$ for liquid).
The energy functional has four separate terms,
\begin{equation} \label{eq:phasefield}
  E[\phi] = E^\mathrm{B}[\phi] + E^\mathrm{I}[\phi] + E^\mathrm{S}[\phi] + E^\mathrm{V}[\phi].
\end{equation}
The first term, $E^\mathrm{B}$, uses a double well potential to set values of $\phi=\pm1$ in the bulk.
The second term then provides the interfacial energy between the liquid and gas by imposing an energy penalty to gradients in $\phi$.
$E^\mathrm{S}$ is the solid-liquid interaction energy, which sets the contact angles of the hydrophilic and hydrophobic regions to 60\si{\degree} and 110\si{\degree}, respectively.
Finally, $E^\mathrm{V}$ constrains the volume of the liquid drop by penalising any variation from the target volume.

These four sections of the phase-field model are obtained using the following equations,
\begin{align}
  E^\mathrm{B}[\phi] &= \sum_i \frac{1}{\epsilon} \left( \frac{{\phi_i}^4}{4} - \frac{{\phi_i}^2}{2} + \frac{1}{4} \right) \upDelta V,\\
  E^\mathrm{I}[\phi] &= \sum_i \frac{\epsilon}{2} \abs{\grad \phi_i}^2 \upDelta V,\\
  E^\mathrm{S}[\phi] &= \sum_j \sqrt{2}\cos\theta_j \left( \frac{{\phi_j}^3}{6} - \frac{\phi_j}{2} - \frac{1}{3} \right) \upDelta S,\\
  E^\mathrm{V}[\phi] &= k_\mathrm{V} \left[ \sum_i \frac{\phi_i + 1}{2} \upDelta V - V_0 \right]^2,
\end{align}
where the index $i$ includes all of the nodes, while $j$ represents the nodes along the solid surface.
$\upDelta V$ and $\upDelta S$ are, respectively, the volume and solid surface areas contained by each individual node.
$\epsilon$ is the liquid-gas interface width (set to 2.5 lattice units), $\theta_j$ is the contact-angle with the solid surface, and $V_0$ is the constrained volume of the liquid drop.
The strength of the volume constraint is parametrised by $k_\mathrm{V}$ for which we use a value of $10^4$.
The gradients of these terms are given by,
\begin{align}
  \frac{\partial E^\mathrm{B}}{\partial \phi_i} &= \frac{1}{\epsilon} \left( {\phi_i}^3 - \phi_i \right) \upDelta V,\\
  \frac{\partial E^\mathrm{I}}{\partial \phi_i} &= \frac{\epsilon}{2} \left[
    \frac{\partial \abs{\grad \phi_i}^2}{\partial \phi_i} +
    \sum_k\frac{\partial \abs{\grad \phi_k}^2}{\partial \phi_i} \right] \upDelta V,\\
  \frac{\partial E^\mathrm{S}}{\partial \phi_j} &= \sqrt{2}\cos\theta_j \left( \frac{{\phi_j}^2}{2} - \frac{1}{2} \right) \upDelta S,\\
  \frac{\partial E^\mathrm{V}}{\partial \phi_i} &= k_\mathrm{V} \left[ \sum_{i'} \frac{\phi_{i'} + 1}{2} \upDelta V - V_0 \right] \upDelta V,
\end{align}
where index $k$ denotes neighbouring nodes at which the evaluation of the gradient uses $\phi_i$.

\section{Implementation details for the CINEB, DNEB-HEVF, and BITSS-HEVF methods}\label{sec:HEVFdetails}
The spring constants used to connect the states in the CINEB and DNEB methods are system dependent.
They are chosen such that they keep the states equidistant without overwhelming the gradients arising from the potential energy landscapes under consideration.
We employ $10^{-1}$ for the Lennard-Jones particle cluster, $10^{-2}$ for the cylindrical buckling example, and $10^{-6}$ for the striped wetting system.

When hybrid eigenvector following is combined with BITSS or DNEB, we need to set out criteria to determine when the double-ended method has sufficiently converged, at which point the hybrid eigenvector following method begins.
For BITSS, the criterion is when the average of the two states changes by less than a tenth of the reduction in the separation given by \cref{eq:diteration} during a BITSS step.
For DNEB, the root-mean-square of the total energy gradient of the chain of states is used with a convergence criterion of $10^{-3}$ for the particle cluster and cylindrical buckling systems, and $10^{-5}$ for the striped wetting.
We employ the hybrid eigenvector following method implemented in the program OPTIM \cite{OPTIM}.

\bibliography{bibexport.bib}

\end{document}


\title{A robust and memory-efficient transition state search method for complex energy landscapes \\ Supplementary Information}
\date{\today}
\author{Samuel J. Avis}
\author{Jack R. Panter}
\email[]{j.r.panter@durham.ac.uk}
\author{Halim Kusumaatmaja}
\email[]{halim.kusumaatmaja@durham.ac.uk}
\affiliation{Department of Physics, Durham University, South Road, Durham DH1 3LE, UK}
\maketitle

\section{Optimising the constraint coefficients}
The constraint coefficients are chosen to set comparable sizes for each term in the gradient of the BITSS potential,
\begin{equation}
  \grad E_\text{BITSS} = \sum_{\substack{i=1,2 \\ j\neq i}} \left[ 1 + 2 \kappa_\text{e} (E_i - E_j) \right] \grad E_i + 2 \kappa_\text{d} (d - d_i) \grad d.
\end{equation}
To obtain the expression for the energy coefficient, $\kappa_\text{e}$, we first assume that the separation is fixed, so the distance term can be ignored.
The coefficient $\kappa_\text{e}$ must be high enough to prevent one state from being pulled over the ridge, for which the greatest risk occurs when the gradient on one state is much greater than the other, e.g. $\abs{\grad E_2} \gg \abs{\grad E_1}$.
In this case the total gradient is approximated by
\begin{equation}
  \grad E_\text{BITSS} = \left[ 1 - 2 \kappa_\text{e} (E_1 - E_2) \right] \grad E_2.
\end{equation}
Therefore, when not at a transition state or a minimum in the landscape ($\abs{\grad E_2} \neq 0$) convergence will occur when the term in the square brackets is zero, resulting in $E_1 - E_2 = 1 / 2 \kappa_\text{e}$.
This energy difference should be less than current energy barrier, so we can substitute it with $E_\text{B} / \alpha$, where $E_\text{B}$ is the energy barrier estimate described in the main paper and $\alpha$ is a constant greater than one.
This leaves us with the expression,
\begin{equation}
  \kappa_\text{e} = \frac {\alpha} {2 E_\text{B}}.
\end{equation}

The distance coefficient is determined by assuming that the energies are equal and thus the energy constraint can be ignored.
In this case, convergence will occur when $\grad E_1 + \grad E_2 = -2 \kappa_\text{d} (d - d_i) \grad d$.
It is then possible to find the value of $\kappa_\text{d}$ for which the magnitude of each side of this equation is equal for a desired relative error in the distance, $\beta = (d - d_i) / d_i$,
\begin{equation}
  \kappa_\text{d} = \frac {\sqrt{\abs{\grad E_1}^2 + \abs{\grad E_2}^2}} {2 \sqrt{2} \beta d_i}.
\end{equation}
(Note: The gradient of the distance with respect to a single state has a magnitude of 1, thus the magnitude of the gradient for the pair of states is $\abs{\grad d} = \sqrt{1^2+1^2} = \sqrt{2}$.)
To ensure that the coefficient is not too small if the gradient is close to zero, such as when the states are initialised at the minima, a lower bound is set by replacing $\abs{\grad E_1}$ and $\abs{\grad E_2}$ with $2 E_\text{B} / d_i$. This gives the expression,
\begin{equation}
  \kappa_\text{d} = \mathrm{max} \! \left(
  \frac {\sqrt{\abs{\grad E_1}^2 + \abs{\grad E_2}^2}} {2 \sqrt{2} \beta d_i}, \quad
  \frac{E_\text{B}}{\beta d_i^2} \right).
\end{equation}

\begin{figure}[htb]
  \includegraphics{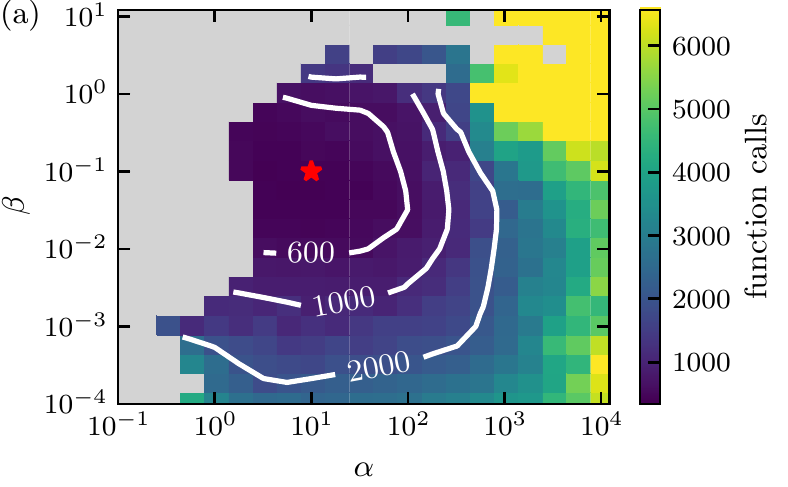}%
  \includegraphics{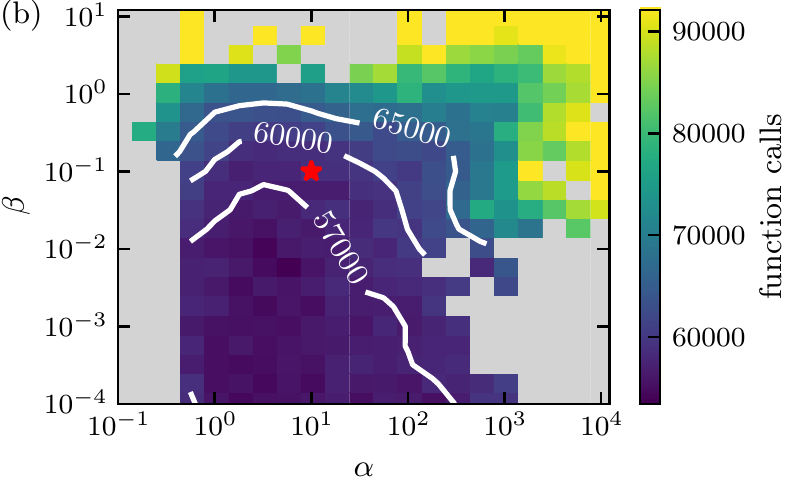}%
  \caption{\label{fig:paramtest}
    Speed of convergence of the BITSS method under different choices of parameters for (a) the seven-particle cluster, and (b) cylindrical buckling.
    The speed is given by the number of evaluations of the gradient until the two states are separated by less than a thousandth of the initial separation.
    The combinations that do not converge to the transition state are shown in grey.
    The chosen parameters are marked by a red star.
  }
\end{figure}

The constant parameters $\alpha$ and $\beta$ still must be chosen.
To this end we have tested different parameter choices using the seven-particle cluster and cylindrical buckling examples, with fractional separation decreases per BITSS step of $f=0.5$ and $f=0.4$, respectively.
\Cref{fig:paramtest} shows which choices lead to convergence and the speed at which this occurs.
The parameters $\alpha = 10$ and $\beta = 0.1$ are chosen for both converging quickly and being situated far from the regions of non-convergence in both test cases.
Hence, this choice is likely to still succeed even if the boudaries of the regions were to shift under different systems.
However, the user is able to choose values optimised for their specific system should they wish.

\section{Multiple transition states}
\begin{figure}[htb]
  \includegraphics{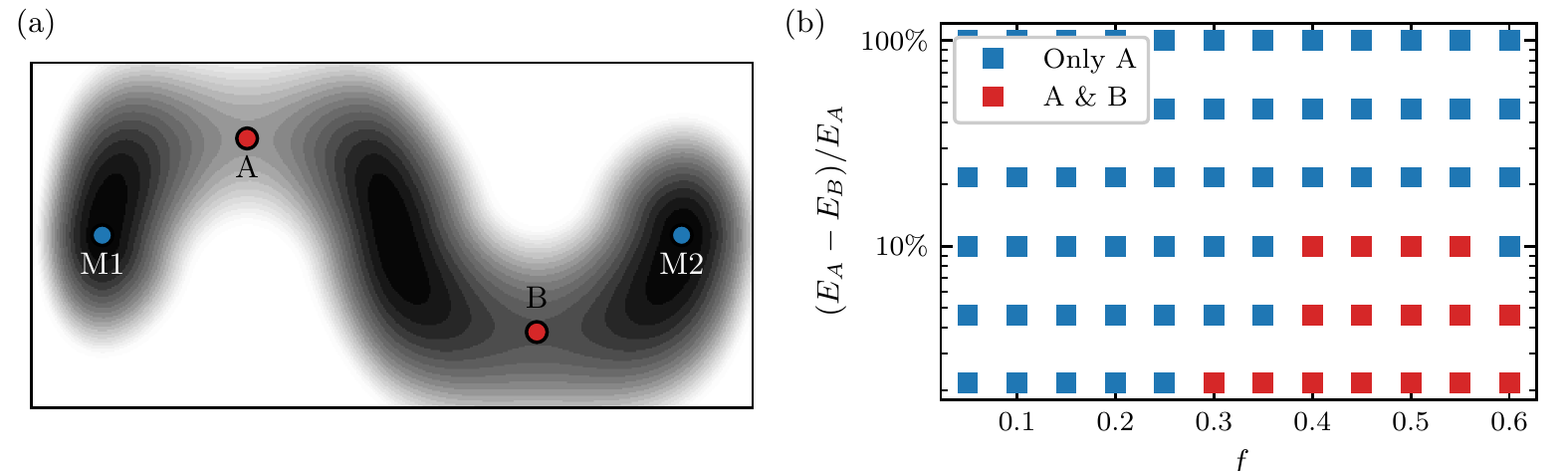}
  \caption{\label{fig:multits}
    (a) The potential used in the test.
    M1 and M2 denote the two starting minima, and A and B are the two transition states.
    The energy of the barrier to B, $E_\text{B}$, is varied between 0 and $E_\text{A}$.
    (b) The results for the parameters under which BITSS reliably converges to A, and the points at which it sometimes converges to B.
    The x-axis is the BITSS distance reduction factor, $f$, and the y-axis is the relative difference between the two barrier heights.
  }
\end{figure}

Here we test how the BITSS method performs when there are multiple transition states in the pathway between the two starting minima.
For this we use a 2D potential, shown in \cref{fig:multits}a, with a pathway that follows a chicane of two $135^\circ$ circular arcs.
The energy is given by the squared distance from this path, plus two barriers resulting in transition states A and B, with energies $E_\text{A}$ and $E_\text{B}$.
We then vary $E_\text{B}$ between 0 and $E_\text{A}$, and the size of the distance reduction factor, $f$, to see if BITSS successfully converges to the higher transition state A.
For each pair of parameters we perform 5 runs with slight variations in the starting positions, with the results in \cref{fig:multits}b showing the points at which all 5 converge to A.

We see that if the difference between the two barriers is sufficiently large ($\gtrsim 10\%$) then BITSS always converges to the higher transition state.
However, as the difference decreases it starts to sometimes converge to the lower transition state if the separation is decreased quickly.
Therefore, to ensure that BITSS always converges to the higher energy transition state it may be necessary to restrict the separation step size.

Although the equal-energy constraint should cause the states to converge to A, this is not always the case because the discrete steps in the minimisation can cause the left image to jump over A before the other image passes B.
If the minisation is fast and takes large steps then the chance for this to occur is increased.
Therefore, systems with complex, high-dimensional landscapes can probably successfully locate the highest transition state for larger values of $f$ and smaller height differences than simpler systems such as this 2D example.

